# Fusion of ECG Foundation Model Embeddings to Improve Early Detection of Acute Coronary Syndromes

Zeyuan MENG [a], Lovely Yeswanth PANCHUMARTHI [a], Saurabh KATARIA [b], Alex FEDOROV [b], Jessica ZÈGRE-HEMSEY [c], Xiao HU [a,b], and Ran XIAO [a,b,1]

[a] *Department of Computer Science, Emory University, USA*
[b] *School of Nursing, Emory University, USA*
[c] *School of Nursing, University of North Carolina at Chapel Hill, USA*

**Abstract.** Acute Coronary Syndrome (ACS) is a life-threatening cardiovascular condition where early and accurate diagnosis is critical for effective treatment and improved patient outcomes. This study explores the use of ECG foundation models, specifically ST-MEM and ECG-FM, to enhance ACS risk assessment using prehospital ECG data collected in the ambulances. Both models leverage self-supervised learning (SSL), with ST-MEM using a reconstruction-based approach and ECG-FM employing contrastive learning, capturing unique spatial and temporal ECG features. We evaluate the performance of these models individually and through a fusion approach, where their embeddings are combined for enhanced prediction. Results demonstrate that both foundation models outperform a baseline ResNet-50 model, with the fusion-based approach achieving the highest performance (AUROC: $0.843 \pm 0.006$, AUCPR: $0.674 \pm 0.012$). These findings highlight the potential of ECG foundation models for early ACS detection and motivate further exploration of advanced fusion strategies to maximize complementary feature utilization.

**Keywords.** electrocardiogram, foundation model, acute coronary syndromes

## 1. Introduction

Acute Coronary Syndrome (ACS) is a life-threatening condition caused by the sudden blockage of coronary arteries(1). Without prompt intervention, ACS can result in myocardial infarction or even death. The electrocardiogram (ECG) remains the gold standard for early risk assessment and triage, with specific ECG changes, such as ST-segment elevation or depression, reflecting coronary artery occlusion (2-4). Early ECG acquisition, often during ambulance transport, facilitates rapid decision-making and reduces "door-to-balloon" time, improving patient outcomes.

    A growing body of research explores the application of machine learning to enhance ECG interpretation, aiming to improve early decision-making in clinical settings. Over time, these efforts have evolved from traditional machine learning techniques, such as decision trees(5), to more advanced deep learning methods(6-8), and more recently, to

---

[1] Corresponding Author: Ran Xiao, email: ran.xiao@emory.edu.

ECG foundation models(9-12). This progression has significantly improved the precision of ACS risk assessment based on ECG data(9, 10). By leveraging self-supervised learning (SSL), ECG foundation models learn comprehensive ECG representations from large datasets without extensive labeled data, which can be both time-consuming and costly. The resultant ECG representations can then be applied to various downstream tasks, reducing the dependency on large, labeled datasets for effective performance. By offloading the feature extraction process during pretraining, these models hold great promise for enhancing ECG-based risk assessment.

The effectiveness of ECG foundation models largely depends on the design of their SSL tasks. For example, ST-MEM employs a reconstruction-based SSL approach using masked 12-lead signals to capture both spatial and temporal features(9), while ECG-FM utilizes contrastive learning with lead-dropping, focusing on spatial relationships across the 12-lead representations of cardiac electrical activity(10). Both models have demonstrated promising performance in detecting ACS(9, 10). Given the differences in SSL strategies and the resultant representations learned by each model, there may be complementary information between representations learned by each model for a given ECG signal. Combining these distinct representations could potentially enhance model performance; however, this hypothesis has yet to be thoroughly evaluated. Furthermore, most evaluations of ECG foundation models for ACS risk assessment have been conducted using hospital-based ECG data, such as those collected in emergency departments. The critical ultra-early stage of care, where ECG data is obtained in ambulances during pre-hospital care, remains largely unexplored. This gap represents a missed opportunity to assess the utility of these models in pre-hospital settings, where early detection can significantly impact patient outcomes.

To address these gaps, our study utilizes pretrained ECG foundation models to assess their performance on an ECG dataset specifically sourced from pre-hospital settings for early ACS detection. We hypothesize that foundation models trained with distinct SSL tasks capture complementary aspects of ECG representations. By integrating embeddings from multiple models, we aim to improve predictive accuracy and develop a more robust solution for early ACS diagnosis and timely intervention in pre-hospital care settings.

## 2. Methods

This study evaluates two ECG foundation models—ST-MEM and ECG-FM—assessing their individual efficacy in ECG representation learning for detecting ACS using prehospital ECG data as the downstream task. To explore the potential complementary effects of their embeddings, the models' outputs are combined in a fusion model to test the hypothesis that integrating multiple ECG representations can enhance performance. These foundation model-based approaches are then compared against a baseline model, which is directly trained and tested on the downstream task data using a supervised learning setup.

### 2.1. Experimental design

**ST-MEM** is pretrained on a composite dataset of 188,480 12-lead ECGs sourced from the Chapman dataset (10,247 ECGs), Ningbo dataset (34,905 ECGs), and CODE-15 dataset (143,328 ECGs)(9). Each recording consists of 10-second ECG signals sampled at either 500 Hz or 400 Hz. The pretraining framework employs a masked autoencoder approach, where patches of the ECG signals are reconstructed. This process enables the model to learn comprehensive ECG representations by capturing both spatial (lead-wise) and temporal information. For the downstream ACS detection task, ST-MEM embeddings for ECG are classified using an XGBoost model with hyperparameters: maximum tree depth of 6, learning rate of 0.1, and subsample/colsample_bytree set to 0.8 for generalization.

**ECG-FM** is pretrained using a contrastive learning approach on a large dataset comprising 1.66 million standard 12-lead ECGs, sourced from UHN-ECG, MIMIC-IV and PhysioNet 2021 Challenge datasets(10). Positive pairs are generated by randomly dropping a lead, while negative pairs use different recordings. The model maximizes agreement between augmented ECG views. For ACS detection, ECG-FM embeddings are processed using the same XGBoost approach as ST-MEM.

In the **fusion approach**, both models' embeddings are min-max normalized, concatenated, and classified using XGBoost with identical hyperparameters. This tests whether combining diverse ECG features improves performance.

A **baseline model**, the standard ResNet-50, processes raw ECG signals from MEDIC dataset using residual blocks and bottleneck designs. It is trained with a learning rate of 0.001, batch size of 64, and for 100 epochs.

*2.2. Downstream task data*

The downstream task data, the MEDIC dataset, were collected through a healthcare registry at Atrium Health (Charlotte, North Carolina) in collaboration with Mecklenburg County Emergency Medical Services(13). The dataset includes patients treated between 2013 and 2017 with suspected ACS who had prehospital ECGs collected by EMS. It consists of 5,813 10-second 12-lead ECG recordings, including 1,207 ACS cases and 4,606 non-ACS cases, resulting in an approximate case ratio of 20% ACS. The data were reshuffled 10 times, with each reshuffling using 80% of the data as the development set for model training and the remaining 20% as an independent test set, ensuring stratification across ACS and non-ACS cases.

*2.3. Performance evaluation*

The area under the receiver operating characteristic curve (AUROC) and the area under the precision-recall curve (AUCPR) were selected as the performance evaluation metrics. AUROC evaluates model performance across all classification thresholds by plotting the true positive rate against the false positive rate. AUCPR focuses on precision and recall, making it effective for imbalanced datasets by highlighting the model's ability to detect positive cases despite class imbalance.

**3. Results**

Figure 1 presents the t-SNE representations of ECG embeddings generated by the two foundation models, ST-MEM (left) and ECG-FM (right), for a qualitative comparison. A total of 500 samples (250 ACS and 250 non-ACS) were randomly selected from the training set of the MEDIC dataset for visualization purposes. The figure reveals distinctly different patterns in the learned representations: ST-MEM produces more compact and clustered ECG embeddings, while ECG-FM generates embeddings that are noticeably more dispersed across the feature space.

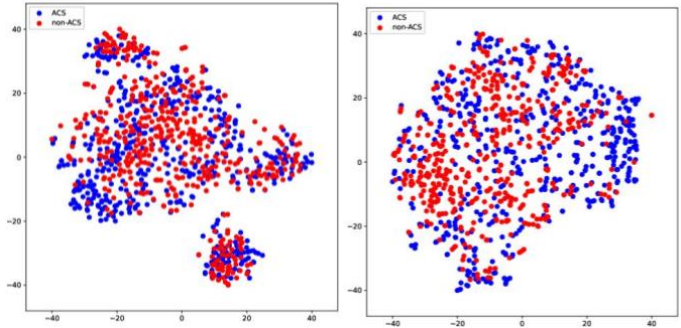

**Figure 1.** t-SNE representations of ECG embeddings from ST-MEM (left) and ECG-FM (right).

Table 1 presents the mean and standard deviation (STD) of model performance across 10 reshufflings for detecting acute coronary syndrome (ACS) using the baseline model and foundation model-based approaches. The results demonstrate that both foundation models (ST-MEM and ECG-FM) consistently outperform the baseline model in terms of both AUROC and AUCPR. Among the foundation models, ECG-FM outperforms ST-MEM across both metrics. Notably, the fusion-based approach, which combines embeddings from both foundation models, achieves the highest performance, yielding an AUROC of $0.843 \pm 0.006$ and an AUCPR of $0.674 \pm 0.012$, indicating further improvements through fusion of embeddings.

**Table 1.** Comparison of model performance in detecting ACS.

|  | **Baseline** | **ST-MEM** | **ECG-FM** | **Fusion** |
|---|---|---|---|---|
| **AUROC** | 0.701±0.471 | 0.783±0.007 | 0.835±0.012 | **0.843±0.006** |
| **AUCPR** | 0.471±0.078 | 0.594±0.013 | 0.667±0.024 | **0.674±0.012** |

## 4. Discussion

The ongoing evolution of AI-driven approaches for enhancing the diagnostic value of ECG is advancing ACS diagnosis toward precision medicine. A key advancement in the field is the increasing use of foundation models, which leverage vast amounts of unlabeled ECG data to improve the model's ability to capture and represent cardiac electrophysiology with greater depth and accuracy—an achievement previously limited by traditional AI methods constrained by labeled data availability. This shift has transferred the feature selection process to the pretraining phase, enabling models to achieve strong performance even when working with limited labeled datasets during downstream tasks. This advantage is evident in our results, where all foundation models tested consistently outperformed the baseline model, highlighting the superior feature extraction capability of foundation models.

Another key finding from our pilot study is the impact of distinct SSL strategies on feature representation across foundation models. ST-MEM employs a reconstruction-based SSL approach, aiming to balance both temporal and spatial aspects of the ECG waveform, while ECG-FM uses a contrastive learning approach, emphasizing lead-level spatial relationships. These differing SSL strategies result in unique representations, as evidenced by the t-SNE plots, where the two models display clearly distinct feature distributions. Our results further suggest that the features learned by these models may be complementary, as the fusion-based approach, which combines embeddings from both models, consistently achieved superior performance across all metrics compared to individual models alone.

While this pilot study implemented fusion through basic embedding concatenation, future research will explore more advanced fusion strategies to further improve performance. Beyond ad hoc feature fusion setups, we aim to investigate hybrid loss functions that integrate model-specific features during the pretraining phase, optimizing the ability to capture both shared and unique ECG representations for enhanced ACS detection.